\documentclass[prm,aps,twocolumn,superscriptaddress,showpacs,nofootinbib]{revtex4-1}

\makeatletter
\renewcommand*{\@fnsymbol}[1]{\ensuremath{\ifcase#1\or \dagger\or \dagger\or \ddagger\or
		\mathsection\or \mathparagraph\or \|\or **\or \dagger\dagger
		\or \ddagger\ddagger \else\@ctrerr\fi}} 

\makeatother
\usepackage{color}
\usepackage{graphicx}
\usepackage{epstopdf}
\usepackage[colorlinks=true,linkcolor=black, citecolor=blue, urlcolor=blue, 
unicode=true,breaklinks=true]{hyperref}

\usepackage{float}
\usepackage{lipsum}
\usepackage{amsmath} 
\usepackage{amssymb}
\usepackage{color}
\usepackage{hyperref} 
\usepackage{anysize}
\usepackage{hhline,multirow,float}
\usepackage{makecell}
\usepackage{array}
\usepackage{dcolumn}
\newcolumntype{?}{!{\vrule width 1pt}}
\usepackage{color}
\usepackage{placeins}
\usepackage{mathtools}

\usepackage{multirow}
\usepackage[table]{xcolor}
\usepackage{capt-of}
\usepackage{wasysym}

\marginsize{1.5 cm}{1.5 cm}{0.5 cm}{0.5 cm}

\usepackage{etoolbox} 
\usepackage{lipsum} 

\makeatletter
\appto{\appendix}{%
	\@ifstar{\def\theequation@prefix{A.}}%
	{}%
}
\makeatother

\begin{document}
\newcommand\bbone{\ensuremath{\mathbbm{1}}}
\newcommand{\ul}{\underline}
\newcommand{\bp}{{\bf p}}
\newcommand{\vl}{v_{_L}}
\newcommand{\vc}{\mathbf}
\newcommand{\be}{\begin{equation}}
\newcommand{\ee}{\end{equation}}
\newcommand{\bk}{{{\bf{k}}}}
\newcommand{\bK}{{{\bf{K}}}}
\newcommand{\cE}{{{\cal E}}}
\newcommand{\bQ}{{{\bf{Q}}}}
\newcommand{\br}{{{\bf{r}}}}
\newcommand{\bg}{{{\bf{g}}}}
\newcommand{\bG}{{{\bf{G}}}}
\newcommand{\hbr}{{\hat{\bf{r}}}}
\newcommand{\bR}{{{\bf{R}}}}
\newcommand{\bq}{{\bf{q}}}
\newcommand{\hx}{{\hat{x}}}
\newcommand{\hy}{{\hat{y}}}
\newcommand{\hd}{{\hat{\delta}}}
\newcommand{\bea}{\begin{eqnarray}}
\newcommand{\eea}{\end{eqnarray}}
\newcommand{\ra}{\rangle}
\newcommand{\la}{\langle}
\renewcommand{\tt}{{\tilde{t}}}
\newcommand{\upa}{\uparrow}
\newcommand{\dna}{\downarrow}
\newcommand{\bS}{{\bf S}}
\newcommand{\vS}{\vec{S}}
\newcommand{\dg}{{\dagger}}
\newcommand{\pdg}{{\phantom\dagger}}
\newcommand{\tphi}{{\tilde\phi}}
\newcommand{\cf}{{\cal F}}
\newcommand{\ca}{{\cal A}}
\renewcommand{\ni}{\noindent}
\newcommand{\ct}{{\cal T}}
\newcommand{\brf}{\bar{F}}
\newcommand{\brg}{\bar{G}}
\newcommand{\jeff}{j_{\rm eff}}
\newcommand{\ngso}{Nd$_{2}$GaSbO$_{7}$}
\makeatletter
\renewcommand*{\@fnsymbol}[1]{\ensuremath{\ifcase#1\or *\or \dagger\or \ddagger\or
   \mathsection\or \mathparagraph\or \|\or **\or \dagger\dagger
   \or \ddagger\ddagger \else\@ctrerr\fi}}
\makeatother

\title{Absence of moment fragmentation in the mixed $B$-site pyrochlore \ngso}

\author{S.~J.~Gomez}\altaffiliation{Both authors contributed equally to this work.}
\affiliation{Materials Department, University of California, Santa Barbara, CA 93106-5050, USA}
\author{P.~M.~Sarte}\altaffiliation{Both authors contributed equally to this work.}
\affiliation{Materials Department, University of California, Santa Barbara, CA 93106-5050, USA}
\affiliation{California NanoSystems Institute, University of California, Santa Barbara, CA 93106-6105, USA}
\affiliation{School of Chemistry, University of Edinburgh, Edinburgh EH9 3FJ, United Kingdom}
\affiliation{Centre for Science at Extreme Conditions, University of Edinburgh, Edinburgh EH9 3FD, United Kingdom}
\author{M.~Zelensky}
\affiliation{Department of Chemistry, University of Winnipeg, Winnipeg, MB R3B 2E9, Canada}
\author{A.~M.~Hallas}
\affiliation{Department of Physics and Astronomy, University of British Columbia, Vancouver, BC V6T 1Z1, Canada}
\affiliation{Stewart Blusson Quantum Matter Institute, University of British Columbia, Vancouver, BC V6T 1Z4, Canada}
\author{B.~A.~Gonzalez}
\affiliation{Materials Department, University of California, Santa Barbara, CA 93106-5050, USA}
\author{K.~H.~Hong}
\affiliation{School of Chemistry, University of Edinburgh, Edinburgh EH9 3FJ, United Kingdom}
\affiliation{Centre for Science at Extreme Conditions, University of Edinburgh, Edinburgh EH9 3FD, United Kingdom}
\author{E.~J.~Pace}
\affiliation{Centre for Science at Extreme Conditions, University of Edinburgh, Edinburgh EH9 3FD, United Kingdom}
\affiliation{School of Physics and Astronomy, University of Edinburgh, Edinburgh EH9 3FJ, United Kingdom}
\author{S.~Calder}
\affiliation{Neutron Scattering Division, Oak Ridge National Laboratory, Oak Ridge, TN 37831, USA}
\author{M.~B.~Stone}
\affiliation{Neutron Scattering Division, Oak Ridge National Laboratory, Oak Ridge, TN 37831, USA}
\author{Y.~Su}
\affiliation{J\"{u}lich Centre for Neutron Science (JCNS) at Heinz Maier-Leibnitz Zentrum (MLZ), Forschungszentrum J\"{u}lich GmbH, Lichtenbergstra{\ss}e 1, 85748 Garching, Germany}
\author{E.~Feng}
\affiliation{J\"{u}lich Centre for Neutron Science (JCNS) at Heinz Maier-Leibnitz Zentrum (MLZ), Forschungszentrum J\"{u}lich GmbH, Lichtenbergstra{\ss}e 1, 85748 Garching, Germany}
\author{M.~D.~Le}
\affiliation{ISIS Facility, Rutherford Appleton Laboratory, Chilton, Didcot OX11 0QX, United Kingdom}
\author{C.~Stock}
\affiliation{Centre for Science at Extreme Conditions, University of Edinburgh, Edinburgh EH9 3FD, United Kingdom}
\affiliation{School of Physics and Astronomy, University of Edinburgh, Edinburgh EH9 3FJ, United Kingdom}
\author{J.~P.~Attfield}
\affiliation{School of Chemistry, University of Edinburgh, Edinburgh EH9 3FJ, United Kingdom}
\affiliation{Centre for Science at Extreme Conditions, University of Edinburgh, Edinburgh EH9 3FD, United Kingdom}
\author{S.~D.~Wilson}
\affiliation{Materials Department, University of California, Santa Barbara, CA 93106-5050, USA}
\affiliation{California NanoSystems Institute, University of California, Santa Barbara, CA 93106-6105, USA}
\author{C.~R.~Wiebe}\altaffiliation{\href{mailto:ch.wiebe@uwinnipeg.ca}{ch.wiebe@uwinnipeg.ca}}
\affiliation{Department of Chemistry, University of Winnipeg, Winnipeg, MB R3B 2E9, Canada}
\affiliation{Department of Chemistry, University of Manitoba, Winnipeg, MB R3T 2N2, Canada}
\affiliation{Department of Physics and Astronomy, McMaster University, Hamilton, ON L8S 4M1, Canada}
\author{A.~A.~Aczel}\altaffiliation{\href{mailto:aczelaa@ornl.gov}{aczelaa@ornl.gov}}
\affiliation{Neutron Scattering Division, Oak Ridge National Laboratory, Oak Ridge, TN 37831, USA}

\date{\today}

\begin{abstract}
Nd-based pyrochlore oxides of the form Nd$_{2}B_{2}$O$_{7}$ have garnered a significant amount of interest owing to the moment fragmentation physics observed in Nd$_{2}$Zr$_{2}$O$_{7}$ and speculated in Nd$_{2}$Hf$_{2}$O$_{7}$. Notably this phenomenon is not ubiquitous in this family, as it is absent in Nd$_{2}$Sn$_{2}$O$_{7}$, which features a smaller ionic radius on the $B$-site. Here, we explore the necessary conditions for moment fragmentation in the Nd pyrochlore family through a detailed study of the mixed $B$-site pyrochlore Nd$_{2}$GaSbO$_{7}$. The $B$-site of this system is characterized by significant disorder and an extremely small average ionic radius. Similarly to Nd$_{2}$Sn$_{2}$O$_{7}$, we find no evidence for moment fragmentation through our bulk characterization and neutron scattering experiments, indicating that chemical pressure (and not necessarily the $B$-site disorder) plays a key role in the presence or absence of this phenomenon in this material family. Surprisingly, the presence of significant $B$-site disorder in Nd$_{2}$GaSbO$_{7}$ does not generate a spin glass ground state and instead the same all-in-all-out magnetic order identified in other Nd pyrochlores is found here.  

\end{abstract}

\maketitle

\FloatBarrier 

\section{Introduction}
Corresponding to a canonical frustrated lattice geometry, the pyrochlore oxides ($A_2B_2$O$_7$) consist of two interpenetrating networks of corner-sharing tetrahedra, composed of the $A$ and $B$-site cations, respectively~\cite{Gardner2010}. In the cases when $A$ and/or $B$ are magnetic, the pyrochlore lattice has played a key role in the realization of a diverse array of exotic magnetic ground states and phenomena that include unconventional long-range order \cite{Oitmaa2013,Petit17:119}, magnetic monopole quasiparticles \cite{Castelnovo2008,Sarte17:29}, classical \cite{Bramwell1998,Ramirez19:399} and quantum \cite{Kimura2013,Ross2011,Zhou2008} spin ices, and other spin liquid states \cite{Gardner1999,Hallas2016}.

A phenomenon that has recently garnered particular interest among these pyrochlores is that of magnetic moment fragmentation, corresponding to the coexistence of magnetic order with a fluctuating spin liquid phase in a single ground state \cite{Brooks-Bartlett2014}. Such fragmentation of the magnetic degrees of freedom arises from the behavior of the magnetic monopole quasiparticles that are elementary excitations of the so-called spin ice phase \cite{01_bramwell}. In this unique state of matter, moments follow local ``two-in-two-out'' ice rules that lead to a macroscopic degeneracy analogous to the degeneracy of configurations in water ice \cite{Harris1997, 99_ramirez}. From the ``two-in-two-out'' picture, flipping a single spin leads to the creation of two magnetic monopoles at the centers of each of the resulting ``3-in-1-out'' and ``1-in-3-out'' tetrahedra. When the monopole density is high, the energy is minimized via the creation of a fragmented state where a crystal of monopoles coexists with a Coulomb spin liquid \cite{Lhotel2020}. Alternatively, this phase can be understood through the superposition of two components: an ordered, divergence-full ``all-in-all-out'' (AIAO) component and a divergence-free ``two-in-two-out'' component which remains fluctuating and provides a gain in entropy \cite{Petit2016}. Experimental realizations of such ``ground-state moment fragmentation'' have been demonstrated in the tripod Kagome materials RE$_{3}$Mg$_{2}$Sb$_{3}$O$_{14}$ (RE = Dy \cite{paddison2016}, Ho~\cite{Dun2020}) and via a staggered magnetic field in the iridate pyrochlores RE$_{2}$Ir$_{2}$O$_{7}$ (RE = Dy~\cite{Cathelin2020}, Ho~\cite{Lefrancois2017}).

In some pyrochlore systems with magnetic $A$-site and non-magnetic $B$-site cations, the fragmentation phenomenon has been proposed to occur in the excitations rather than the ground state \cite{Benton2016}. This ``excited-state moment fragmentation" has been identified in Nd$_{2}$Zr$_{2}$O$_{7}$ via neutron scattering measurements \cite{Petit2016}. Two magnetic excitations are observed in this data: a nearly-dispersionless mode and a dispersive mode corresponding to the divergence-free and divergence-full components of the fragmentation, respectively. The former assignment was confirmed via the observation of pinch points in the $Q$-dependence of the dispersionless mode. This type of fragmentation typically requires a net ferromagnetic interaction between nearest neighbor spins on the pyrochlore lattice and dipolar-octupolar symmetry of the ground state wavefunction doublet, which allows for dipolar ordering of the $J_{z}$ component of the moment while the octupolar $J_{\pm}$ component enables persistent fluctuations \cite{Petit2016,Benton2016}. Nd$_{2}$Hf$_{2}$O$_{7}$ \cite{Anand2015} has been proposed as an additional candidate for realizing this exotic phenomenon. On the other hand, recent work on Nd$_2$Sn$_2$O$_7$ has argued that excited-state moment fragmentation is not realized in this material due to the overall net antiferromagnetic interactions in the ground state, causing moments to fully order into an AIAO structure~\cite{Bertin2015}.

Motivated by the potential role that disorder plays in the realization of quantum spin liquids~\cite{Martin17:7} and other exotic magnetic phenomena including moment fragmentation, there has been a renewed interest in mixed-site pyrochlores \cite{Reig-I-Plessis2020}. Notable examples include the mixed $A$-site fluoride pyrochlore NaCaCo$_2$F$_7$~\cite{Ross2017}, which is argued to form a disorder-induced cluster spin glass phase \cite{Sarkar2017}, and the related pyrochlore NaCaNi$_2$F$_7$ \cite{Plumb2019}, which exhibits disorder-induced spin freezing that appears to preempt a Heisenberg spin liquid state. In addition, the mixed $B$-site oxide pyrochlore Yb$_2$GaSbO$_7$~\cite{Hodges11:23} is a candidate for disorder-induced spin liquid or spin-singlet behavior~\cite{21_sarte}. Interestingly, Nd$_2$ScNbO$_7$ shows the combination of AIAO long-range magnetic order and diffuse scattering expected for moment fragmentation, although open questions remain about the true nature of its magnetic ground state and the presence of this exotic phenomenon in this material \cite{mauws2019, 21_scheie}. More specifically, the diffuse scattering manifests as rods along the $\langle111\rangle$ directions, but no evidence for pinch points is observed, possibly due to the significant $B$-site disorder in the system. Furthermore, the polarized neutron scattering experiment was not energy-resolved, so it is not clear if the diffuse scattering is in the elastic or inelastic channel. Despite these issues, the previous work on Nd$_2$ScNbO$_7$ indicates that $B$-site disorder may be playing an intriguing role in governing the magnetic properties of this system. On the other hand, the properties of Nd$_2$ScNbO$_7$ may be largely dictated by its average $B$-site ionic radius, which is intermediate between the values for moment fragmentation systems (i.e., Nd$_{2}$Zr$_{2}$O$_{7}$ and Nd$_{2}$Hf$_{2}$O$_{7}$) and the conventional AIAO ordered magnet Nd$_{2}$Sn$_{2}$O$_{7}$. 

The current work aims to disentangle the contributions of $B$-site disorder and chemical pressure to the magnetic properties of the Nd pyrochlores by characterizing the mixed $B$-site pyrochlore Nd$_2$GaSbO$_7$ in detail. Although the expected level of $B$-site disorder is similar to the amount in Nd$_2$ScNbO$_7$, the average $B$-site ionic radius is smaller as compared to all other Nd pyrochlores previously studied. Through a combination of bulk characterization and neutron scattering experiments, we identify AIAO long range antiferromagnetic order in \ngso~below $T{\rm{_N}}=1.1~{\rm{K}}$ with no evidence for moment fragmentation. Our work suggests that chemical pressure plays a key role in establishing excited-state moment fragmentation in Nd pyrochlores, while significant $B$-site disorder has minimal impact on the collective magnetic properties of these systems.

\FloatBarrier 

\section{Experimental Details}

Polycrystalline samples of \ngso~and a non-magnetic mixed $B$-site lattice standard analog Lu$_{2}$GaSbO$_{7}$ were both synthesized by a standard solid state reaction of RE$_{2}$O$_{3}$ (RE = Nd$^{3+}$ or Lu$^{3+}$), Ga$_{2}$O$_{3}$, and Sb$_{2}$O$_{5}$, as previously reported by Strobel \emph{et al.}~\cite{Strobel2010}. Rietveld refinement of fits to the room temperature powder diffraction patterns (Bruker D2 Phaser, Cu K$_{\alpha}$) confirmed the presence of single phase pure RE$_{2}$GaSbO$_{7}$ (RE = Nd$^{3+}$ or Lu$^{3+}$), possessing $Fd\bar{3}m$ symmetry with no discernible impurities ($< 5$\%).

The field and temperature dependence of the DC magnetization of polycrystalline \ngso~were first measured using a Quantum Design 9~T Dynacool Physical Property Measurement System (PPMS) employing the vibrating sample magnetometer (VSM) option with a base temperature of 1.8~K. Subsequent DC magnetization measurements employed the traditional DC scan mode on the Quantum Design Magnetic Property Measurement System (MPMS) equipped with a $^{3}$He insert, providing a base temperature of 420~mK.  Specific heat measurements of polycrystalline \ngso~and its corresponding lattice analog Lu$_{2}$GaSbO$_{7}$ were performed with a Quantum Design 9~T PPMS equipped with a $^{3}\rm{He}$ insert, providing a base temperature of 350~mK.

High-energy inelastic neutron scattering (INS) measurements were performed on the direct-geometry time-of-flight chopper spectrometer SEQUOIA of the Spallation Neutron Source (SNS) at Oak Ridge National Laboratory (ORNL). Approximately 5~g of \ngso~powder was loaded in a cylindrical Al can and cooled to a base temperature of 5~K using a closed cycle refrigerator (CCR). Powder-averaged intensity spectra as a function of momentum transfer $Q$ and energy transfer $E$ were collected with incident energies $E_{i}$ of 60~meV and 150~meV. A fine-resolution Fermi chopper was operated at a frequency of 420~Hz and 600~Hz, respectively, while the $t_{0}$ chopper was spun at 90~Hz, yielding an elastic energy resolution $\frac{\Delta E}{E_{i}} \sim$~2\% \cite{GRANROTH20061104}. As an estimate for a background contribution, an empty Al can was measured in identical experimental conditions for approximately one fifth of the sample counting time. Normalization with a vanadium standard was performed to account for variations of the detector response and the solid angle coverage.

Neutron powder diffraction (NPD) was performed on the high-resolution powder diffractometer HB-2A of the High Flux Isotope Reactor (HFIR) at ORNL \cite{Calder2018}. A vertically focussing Ge(113) wafer-stack monochromator provided a neutron wavelength of 2.41~\AA. The collimation settings were open-21$'$-12$'$ for the premonochromator, monochromator-sample and sample-detector positions, respectively, providing a resolution of $\frac{\Delta d}{d} \sim 2 \times 10^{-3}$. Approximately 5~g of \ngso~powder was pressed into pellets and loaded in a cylindrical Al can with a Cu lid. In the case of the initial 0~T diffractograms, a $^{3}$He insert provided a base temperature of 475~mK. For subsequent measurements in the presence of magnetic fields, the sample can was loaded into a symmetric vertical field cryomagnet which provided magnetic fields up to 5~T and a base temperature of 1.5~K. Rietveld refinements were performed with \texttt{TOPAS Academic V6}~\cite{TOPAS} and the magnetic structure symmetry analysis was performed using \texttt{TOPAS}, \texttt{ISODISTORT}~\cite{Isodistort2}, and \texttt{SARAh}~\cite{Wills2000}.

Low energy INS measurements were performed on the indirect-geometry time-of-flight spectrometer IRIS at the ISIS spallation source. As an indirect-geometry spectrometer, the final energy $E_{f}$ was fixed at 1.84~meV by cooled pyrolytic graphite PG002 analyzer crystals in near-backscattering geometry, providing a range in momentum transfer $Q$ of 0.44~\AA$^{-1}$ to 1.85~\AA$^{-1}$ at the elastic line. The graphite analyzers were cooled to reduce thermal diffuse scattering, providing an elastic resolution of 17.5~$\mu$eV. To optimize incident neutron flux, the \texttt{PG002\textunderscore OFFSET} configuration was chosen where the two disk choppers are spun at 50~Hz, providing an energy transfer window of $-$0.3~meV to 1.2~meV. Approximately 5~g of \ngso~powder was loaded into an annular Cu can and then placed in an Oxford Instruments KelvinoxVT $^{3}$He/$^{4}$He dilution refrigerator (DR) insert of a cryostat, providing a base temperature of 40~mK. Normalization with a vanadium standard was performed to account for variations of the detector response and the solid angle coverage.

\begin{figure*}[htb!]
	\centering
	\includegraphics[width=1\linewidth]{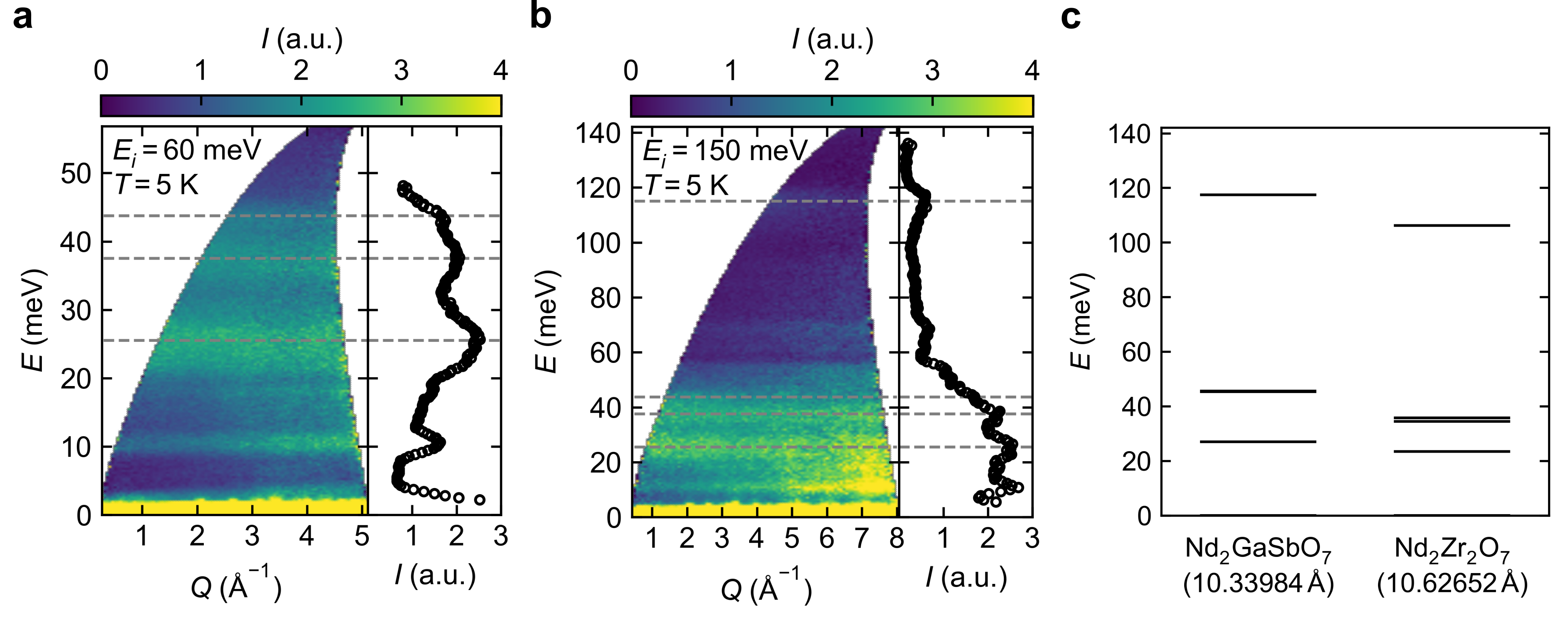}
	\caption{Powder-averaged neutron scattering intensity maps of polycrystalline \ngso~measured on SEQUOIA at 5~K using incident energies (a) $E_i =$~60~${\rm{meV}}$ and (b) $E_i =$~150~${\rm{meV}}$. The corresponding $Q$-cuts have integration ranges of $Q =$~[1, 3]~${\rm{\AA}}^{-1}$ and [3, 6]~${\rm{\AA}}^{-1}$, respectively). Dashed lines indicated approximate energies of the observed excitations. (c) Calculated energy eigenvalues determined from a scaling analysis on Nd$_2$Zr$_2$O$_7$~\cite{Hatnean2015}, exhibiting good agreement with the experimental spectra.}
    \label{figure:sequoia}
\end{figure*}

Polarized diffuse neutron scattering experiments were performed on the diffuse scattering cold neutron time-of-flight spectrometer DNS at the Heinz Maier-Leibnitz Zentrum (MLZ) with an incident neutron wavelength of 4.2~\AA~selected with a double-focusing pyrolytic graphite monochromator. Data was collected in non-time-of-flight mode, leading to the extraction of the integrated scattering intensity with energy transfers up to an $E_i =$~4.64~meV. Approximately 5~g of \ngso~powder was placed in an annular cylindrical can made with oxygen-free Cu and sealed in a He atmosphere. A small amount of deuterated methanol (Sigma-Aldrich, $\geq$ 99.8 at.~\% D) was added to optimize sample-cryostat thermal contact. The Cu can was loaded onto a dry-type DR insert of a top-loading CCR, providing a base temperature of 83~mK. A vanadium standard was used to account for variations of the detector response and the solid angle coverage. Corrections for polarization efficiency of the supermirror analyzers were made by using the scattering from a NiCr alloy. Equal counting times were spent on measuring the scattering along the $x$, $y$, and $z$ directions. The nuclear, magnetic, and nuclear spin-incoherent scattering cross sections were separated as a function of $Q$ using the 6 pt. $xyz$-polarization analysis method \cite{93_scharpf}. The non-spin-flip (NSF) and spin-flip (SF) scattering along each of the three directions was measured with a time ratio of 1:1.

\section{Results and Discussion}

\subsection{Single-Ion Physics}

High-energy inelastic neutron scattering measurements on SEQUOIA at 5~K, shown in Fig.~\ref{figure:sequoia}, identified multiple excitations consistent with crystal field transitions centered at energy transfers of 26, 38, 44, and 115 meV. Although a fifth excitation centered at 11 meV also has significant intensity at low-$Q$, the intensity of this mode increases quickly with $Q$ and therefore we attribute it the same phonon mode observed in other Nd pyrochlores \cite{Anand2017}. The four magnetic excitations identified correspond to the maximum number of levels expected for the $J =$~9/2 multiplet of the Nd$^{3+}$ ions in a crystal field environment. While the data confirmed the presence of a thermally isolated ground state doublet, the $B$-site mixing inherent to \ngso~has broadened the crystal field excitations significantly, which is consistent with a distribution of local crystalline environments for the Nd$^{3+}$ ions. Such broadening has been previously observed in other mixed $B$-site pyrochlores including Tb$_{2}$Sn$_{2-x}$Ti$_{x}$O$_{7}$~\cite{Gaulin2015} and Yb$_{2}$GaSbO$_{7}$~\cite{21_sarte}, transforming the determination of crystal field parameters, even average ones, into a severely under-constrained problem for conventional approaches. 

Instead, as was employed for its ytterbium analog, a quantitative estimate of the Stevens parameters $B^m_l$ comprising the crystal field Hamiltonian $\hat{\mathcal{H}}_{\rm{CEF}}$ for Nd$^{3+}$ with local $D_{3d}$ point group symmetry given by 
\begin{equation}
\begin{split} \label{eq:cefhamiltonian} 
    \hat{\mathcal{H}}_{\rm{CEF}} =& B^0_2\hat{\mathcal{O}}^0_2 + B^0_4\hat{\mathcal{O}}^0_4 + B^3_4\hat{\mathcal{O}}^3_4 + B^0_6\hat{\mathcal{O}}^0_6\\
    &+ B^3_6\hat{\mathcal{O}}^3_6 + B^6_6\hat{\mathcal{O}}^6_6,
\end{split}
\end{equation}\\
was determined using a scaling analysis procedure that has been employed successfully for many other pyrochlore systems~\cite{Bertin2012}. 

\begin{figure*}[t!]
	\centering
	\includegraphics[width=1\linewidth]{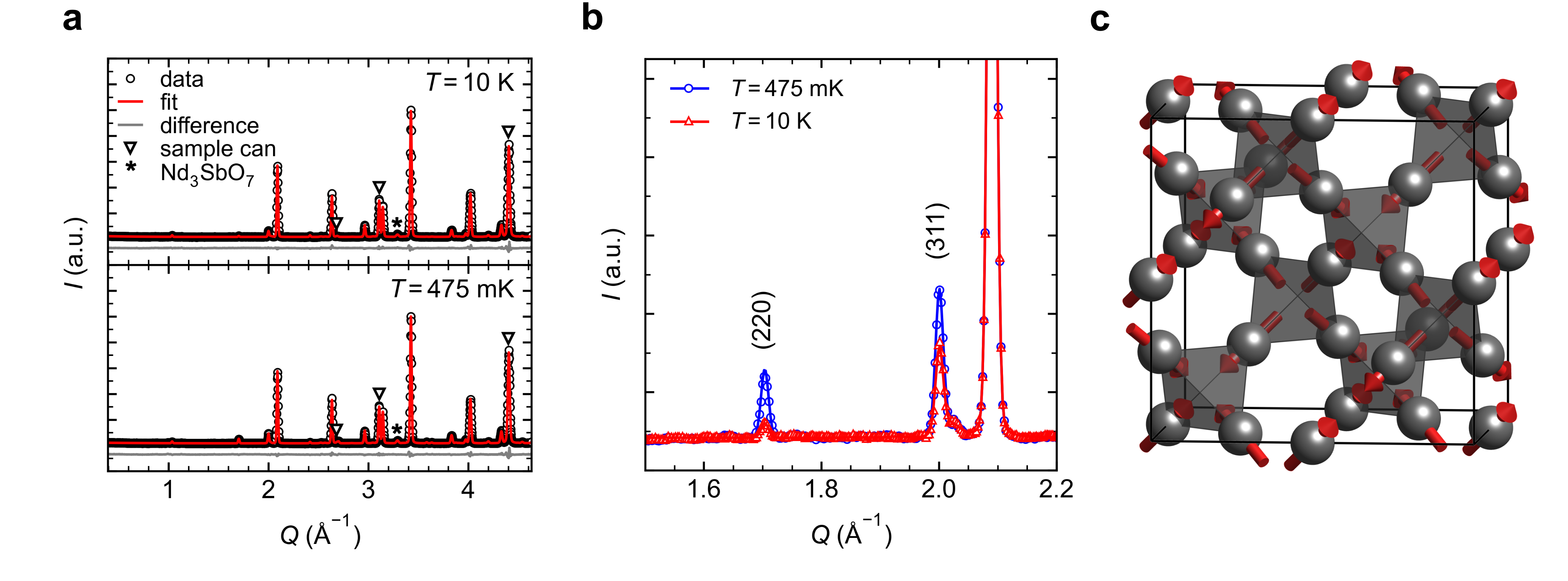}
	\caption{(a) Rietveld fits to neutron powder diffraction profiles for polycrystalline \ngso~collected on HB-2A at 10~K and 475~mK. A small fluorite-related Nd$_3$SbO$_7$ magnetic impurity ($<4\%$) is present. Contributions from the Bragg reflections of the sample can were fit using a Pawley refinement. (b) Enlarged version of the neutron powder diffraction data presented in (a), shown over a limited $Q$ range only. Additional scattering with a magnetic origin appears below $T_{\rm{N}}$ at the (220) and (311) Bragg peak positions. (c) A schematic of the all-in-all-out antiferromagnetic structure that is realized for \ngso~below $T_{\rm{N}}$. To improve clarity, only the Nd$^{3+}$ ions are shown.}
	\label{fig:hb2a}
\end{figure*}

Estimates for the six Stevens parameters of \ngso~(NGSO) were determined by scaling their respective values reported for Nd$_{2}$Zr$_{2}$O$_{7}$ (NZO) according to the relation
\begin{equation} \label{eq:scaling}
B^m_l({\rm{NGSO}}) = \frac{a^{l+1}({\rm{NZO}})}{a^{l+1}({\rm{NGSO}})} B^m_l({\rm{NZO}}),
\end{equation}
where $a$ denotes the lattice parameter for each compound. Lattice parameters of 10.62652(9)~\AA~and 10.33984(6)~\AA~were used for Nd$_{2}$Zr$_{2}$O$_{7}$~\cite{Hatnean2015} and \ngso~respectively. The latter corresponds to the refined lattice parameter obtained from the neutron diffraction data shown in Fig.~\ref{fig:hb2a}.

\begin{table}[tb!]
\caption{\label{table:scaling}
Determination of the Stevens parameters for \ngso~from the corresponding parameters reported for Nd$_{2}$Zr$_{2}$O$_{7}$~\cite{Hatnean2015} using the scaling relation given by Eq.~\ref{eq:scaling}~\cite{Bertin2012}.
}
\begin{ruledtabular}
\begin{tabular}{cccccc}
$B^{m}_{l}$ & $m$ & $l$ & Nd$_{2}$Zr$_{2}$O$_{7}$ & Scaling Factor & \ngso\\ 
\colrule
        $B^0_2$ & $0$ & $2$ & $-$0.179     & 1.08550 & $-$0.199\\
        $B^0_4$ & $0$ & $4$ & $-$0.01399   & 1.14653 & $-$0.01655\\
        $B^3_4$ & $3$ & $4$ & $-$0.1121    & 1.14653 & $-$0.1386\\
        $B^0_6$ & $0$ & $6$ & $-$0.0002944 & 1.12110 & $-$0.0003536\\
        $B^3_6$ & $3$ & $6$ & 0.00358      & 1.12110 & 0.00434\\
        $B^6_6$ & $6$ & $6$ & $-$0.00403   & 1.12110 & $-$0.00553\\
\end{tabular}
\end{ruledtabular}
\end{table}

\begin{table*}[ht!]
 \caption{Energy eigenvalues (in meV) and eigenvectors (in $|J=\frac{9}{2}, m_{J} \rangle$ basis) of $\hat{\mathcal{H}}_{\rm{CEF}}$ for \ngso~employing the six scaled Stevens parameters that are summarized in Table~\ref{table:scaling}.} 
    \label{table:eigensystem}
    \scalebox{1.0}{
    \begin{ruledtabular}
        \begin{tabular}{ccccccccccc}
        $E$~(meV) & $|-\frac{9}{2}\rangle$ & $|-\frac{7}{2}\rangle$ & $|-\frac{5}{2}\rangle$ & $|-\frac{3}{2}\rangle$ & $|-\frac{1}{2}\rangle$ & $|\frac{1}{2}\rangle$ & $|\frac{3}{2}\rangle$ & $|\frac{5}{2}\rangle$ & $|\frac{7}{2}\rangle$ & $|\frac{9}{2}\rangle$\\ [0.5ex]
        \colrule
         0.0     & 0.89     & 0       & 0       & 0.07    & 0       & 0       & 0.45    & 0       & 0        & 0       \\
         0.0     & 0        & 0       & 0       & 0.45    & 0       & 0       & $-$0.07 & 0       & 0        & 0.89    \\
         27.0    & 0        & 0       & 0.61    & 0       & $-$0.01 & $-$0.79 & 0       & $-$0.01 & 0.04     & 0       \\
         27.0    & 0        & 0.04    & 0.01    & 0       & 0.79    & $-$0.01 & 0       & 0.61    & 0        & 0       \\
         45.3    & $-$0.46  & 0       & 0       & 0.14    & 0       & 0       & 0.88    & 0       & 0        & 0       \\
         45.3    & 0        & 0       & 0       & 0.88    & 0       & 0       & $-$0.14 & 0       & 0        & $-$0.46 \\
         45.6    & 0        & 0.55    & 0.09    & 0       & $-$0.51 & 0.07    & 0       & 0.64    & 0.08     & 0       \\
         45.6    & 0        & 0.08    & $-$0.64 & 0       & $-$0.07 & $-$0.51 & 0       & 0.09    & $-$0.55  & 0       \\
         117.4   & 0        & $-$0.83 & 0       & 0       & $-$0.32 & 0       & 0       & 0.46    & 0        & 0       \\
         117.4   & 0        & 0       & 0.46    & 0       & 0       & 0.32    & 0       & 0       & $-$0.83  & 0       \\
        \end{tabular}
    \end{ruledtabular}}
\end{table*}
\indent Employing the six scaled Stevens parameters for \ngso~summarized in Table~\ref{table:scaling}, the crystal field Hamiltonian (Eq.~\ref{eq:cefhamiltonian}) yields four excited doublets, consistent with Kramers' theorem for the $^4I_{9/2}$ ground state multiplet of the 4f$^{3}$ Nd$^{3+}$ ions. As illustrated in Fig.~\ref{figure:sequoia}(c), the energy eigenvalues predicted by the scaling analysis summarized in Table~\ref{table:eigensystem} are in good agreement with the broad crystal field excitations identified above. A closer inspection of the ground state doublet eigenvectors identified a dipolar-octupolar wavefunction of the form $\alpha |\pm \frac{3}{2} \rangle + \beta | \mp \frac{9}{2} \rangle$, where changes in angular momentum are restricted by symmetry to units of three. The coefficients $\alpha < \beta$ suggest that the moments have an Ising-like nature.

The components of the $g$-tensor for the ground state doublet can be calculated according to:
\begin{equation}
    g_z = 2 g_J \left| \langle 0^{\pm} | J_z | 0^{\pm} \rangle \right|,
   \label{eq:gz}
\end{equation}
\begin{equation}
    g_{xy} = 2 g_J \left| \langle 0^{\pm} | J_x | 0^{\mp} \rangle \right| = 2 g_J \left| \pm i \langle 0^{\pm} | J_y | 0^{\mp} \rangle \right|,
    \label{eq:gperp}
\end{equation}
where $g_J = \frac{8}{11}$ corresponds to the Land\'e $g$-factor for the $^4I_{9/2}$ ground state multiplet of a free Nd$^{3+}$ ion, $|0 ^\pm \rangle$ are the eigenvectors of the ground state doublet, and $J_x$, $J_y$, and $J_z$ are total angular momentum operators. We find $g_z$ and $g_{xy}$ values corresponding to 4.73 and 0 respectively, which are indicative of an Ising system and an isotropic $g$-factor:
\begin{equation} \label{eq:isotropic}
    \tilde{g} = \sqrt{\frac{1}{3} (g_z^2 + 2 g_{xy}^2)}
\end{equation}
equal to 2.73.

\subsection{Bulk Property Characterization}

\begin{figure*}[htb!]
	\centering
	\includegraphics[width=1.0\linewidth]{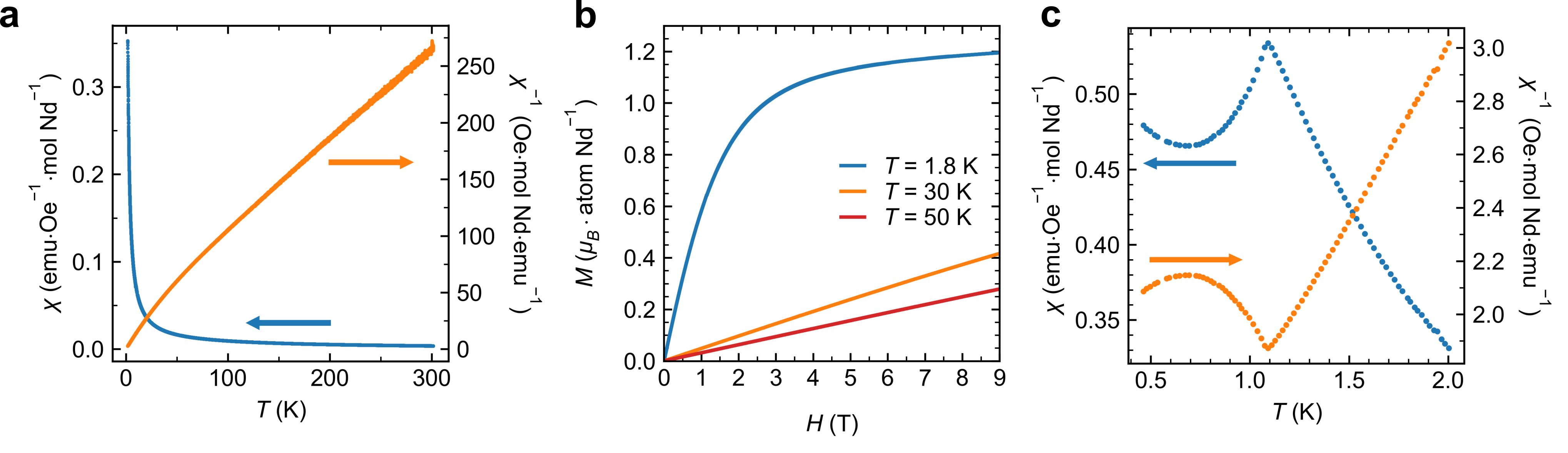}
	\caption{(a) Temperature dependence of the zero field cooled (ZFC) magnetic susceptibility of polycrystalline \ngso~in an external field of $\mu_{0}H =$~0.1~T. A Curie-Weiss fit (dashed line) from 1.8~K to 15~K yields a Curie-Weiss temperature $\theta_{\rm{CW}}=-0.35(1)~\rm{K}$, and an effective paramagnetic moment $\mu_{\rm{eff}}=2.415(1)~\mu_{\rm{B}}$/Nd$^{3+}$. (b) Field-dependence of the isothermal magnetization of polycrystalline \ngso~for various temperatures. At 1.8~K, the magnetization saturates at $\mu_{\rm{sat}} \sim 1.2~\mu_{\rm{B}}/\rm{Nd^{3+}}$. (c) Temperature dependence of the ZFC magnetic susceptibility of polycrystalline \ngso~in an external field of $\mu_{0}H =$~0.1~T extended into the $^{3}$He regime reveals a sharp transition at $T{\rm{_{N}}}$=1.1~K, corresponding to the onset of long range antiferromagnetic order. }
	\label{fig:XvTandMvH}
\end{figure*}

An initial investigation of the collective magnetic ground state of \ngso~was achieved through a combination of DC magnetization and heat capacity measurements. As illustrated in Fig.~\ref{fig:XvTandMvH}(a), the temperature dependence of the DC magnetic susceptibility ($\mu_{0}H =$~0.1~T) revealed an absence of long range magnetic order down to 1.8~K, consistent with the reported behavior for all Nd-based pyrochlores. In the high temperature limit, the inverse susceptibility exhibits linear Curie-Weiss behavior given by 
\begin{equation}
\chi(T) = \frac{C}{T-\theta{\rm{_{CW}}}}.    
\label{eq:CW}
\end{equation}
Over the temperature range $100~{\rm{K}}\leq T \leq 250~{\rm{K}}$, a Curie-Weiss fit yields a Curie-Weiss temperature of $\theta{\rm{_{CW}}} =$~-32.32(7)~K and a Curie constant of $C=1.2481(5)~{\rm{emu \cdot K/Oe \cdot mol~Nd^{3+}}}$, corresponding to an effective paramagnetic moment $\mu{\rm{_{eff}}} =$~3.160(1)~$\mu{\rm{_{B}}}$/Nd$^{3+}$, which is a value in close agreement with the free Nd$^{3+}$ moment of 3.62~$\mu{\rm{_{B}}}$.

As the temperature decreases, the inverse susceptibility deviates significantly from the high-temperature Curie-Weiss behavior due to crystal field effects on the $^4I_{9/2}$ ground state multiplet~\cite{hallas15:91}. As a result, characterization of the crystal field ground state through the Curie-Weiss law was limited to low temperatures. Employing a temperature range of $1.8~{\rm{K}}\leq T \leq 15~{\rm{K}}$, a fit to the Curie-Weiss law (Eq.~\ref{eq:CW}) yields a negative $\theta_{\rm{CW}} =$~-0.35(1)~$\rm{K}$ and a Curie constant $C = $~0.729(1)~${\rm{emu \cdot K/Oe \cdot mol~Nd^{3+}}}$, corresponding to a $\mu_{\rm{eff}} =~2.415(1)~\mu_{\rm{B}}/\rm{Nd}^{3+}$. Both Curie-Weiss parameters are quantitatively very similar to those reported for Nd$_{2}$Sn$_{2}$O$_{7}$~\cite{bondah01:79,Bertin2015}. The negative Curie-Weiss temperature is particularly noteworthy, as the moment fragmentation systems Nd$_{2}$Zr$_{2}$O$_{7}$ \cite{Xu2015, Lhotel2015} and Nd$_{2}$Hf$_{2}$O$_{7}$ \cite{Anand2015} have positive values for this parameter.

The effective moment for \ngso~is consistent with a thermally isolated ground state doublet that is not purely $|J=9/2,m_J=\pm 9/2\rangle$ in character, which is in good agreement with the CEF scaling analysis results described above. The ground state wavefunctions may even contain some contributions from the excited $^{4}I_{11/2}$ multiplet, as has been observed in Nd$_{2}$Zr$_{2}$O$_{7}$ \cite{Xu2015}. By exploiting the presence of a large gap $\Delta\sim$20~meV between the crystal field ground state and first excited state previously identified in Fig.~\ref{figure:sequoia}, a $J{\rm{_{eff}}}=\frac{1}{2}$ model can be employed to account for the experimental data. In the case of this simplified model, the $J$ multiplet with the Land\'{e} $g$-factor $g_{J}$ is projected onto a $J\rm{_{eff}}=\frac{1}{2}$ manifold with an isotropic $\tilde{g}$ factor defined by Eq.~\ref{eq:isotropic} with components $g_{z}$ and $g_{xy}$ that are given by Eqs.~\ref{eq:gz} and~\ref{eq:gperp}, respectively. By employing such a projection, the effective paramagnetic moment of $g_{J}\sqrt{J(J+1)}$ for a polycrystalline sample is simply redefined in terms of the projected values of $\tilde{g}$ and $J\rm{_{eff}}$ as~\cite{sarte18:98,sarte18:98_2}         
\begin{equation} \label{eq:new}
    \mu_{\rm{eff}} = \tilde{g}\sqrt{J_{\rm{eff}}(J_{\rm{eff}}+1)}~\mu_{\rm{B}}.
\end{equation}
For \ngso, the effective spin-1/2 model with the $\tilde{g}$ value taken from the crystal field scaling analysis described above yields $\mu_{\rm{eff}} = 2.37(1)~\mu_{\rm{B}}$, in excellent agreement with the Curie-Weiss law fitted value in the low temperature limit.
 
The field dependence of the isothermal magnetization at selected temperatures, illustrated in Fig.~\ref{fig:XvTandMvH}(b), provides additional support for the applicability of the $J{\rm{_{eff}}}=\frac{1}{2}$ model in \ngso~and the realization of local $\langle 111 \rangle$ Ising anisotropy for the Nd$^{3+}$ spins. The saturation magnetization values along the different high-symmetry directions in this case \cite{Xu2015} are given by:
\begin{equation}
    M_{\rm{sat}}^{100} = \frac{1}{\sqrt{3}}\mu_{\rm{CEF}}, 
\end{equation}
\begin{equation}
    M_{\rm{sat}}^{110} = \frac{1}{\sqrt{6}}\mu_{\rm{CEF}}, 
\end{equation}
\begin{equation}
    M_{\rm{sat}}^{111} = \frac{1}{2}\mu_{\rm{CEF}},
\end{equation}
and the powder-averaged saturation magnetization is:
\begin{equation}
    M_{\rm{sat}}^{p} = \frac{6M_{\rm{sat}}^{100} + 12M_{\rm{sat}}^{110} + 8M_{\rm{sat}}^{111}}{26}.
\end{equation}
The measured magnetization of 1.2~$\mu_B$ at 9~T and 1.8~K agrees remarkably well with the calculated value of 1.13~$\mu_B$ determined from the crystal field scaling analysis results described above, taking $\mu_{\rm{CEF}} = \mu_{\rm{eff}}$.

\begin{figure}[htb!]
	\centering
	\includegraphics[width=0.9\linewidth]{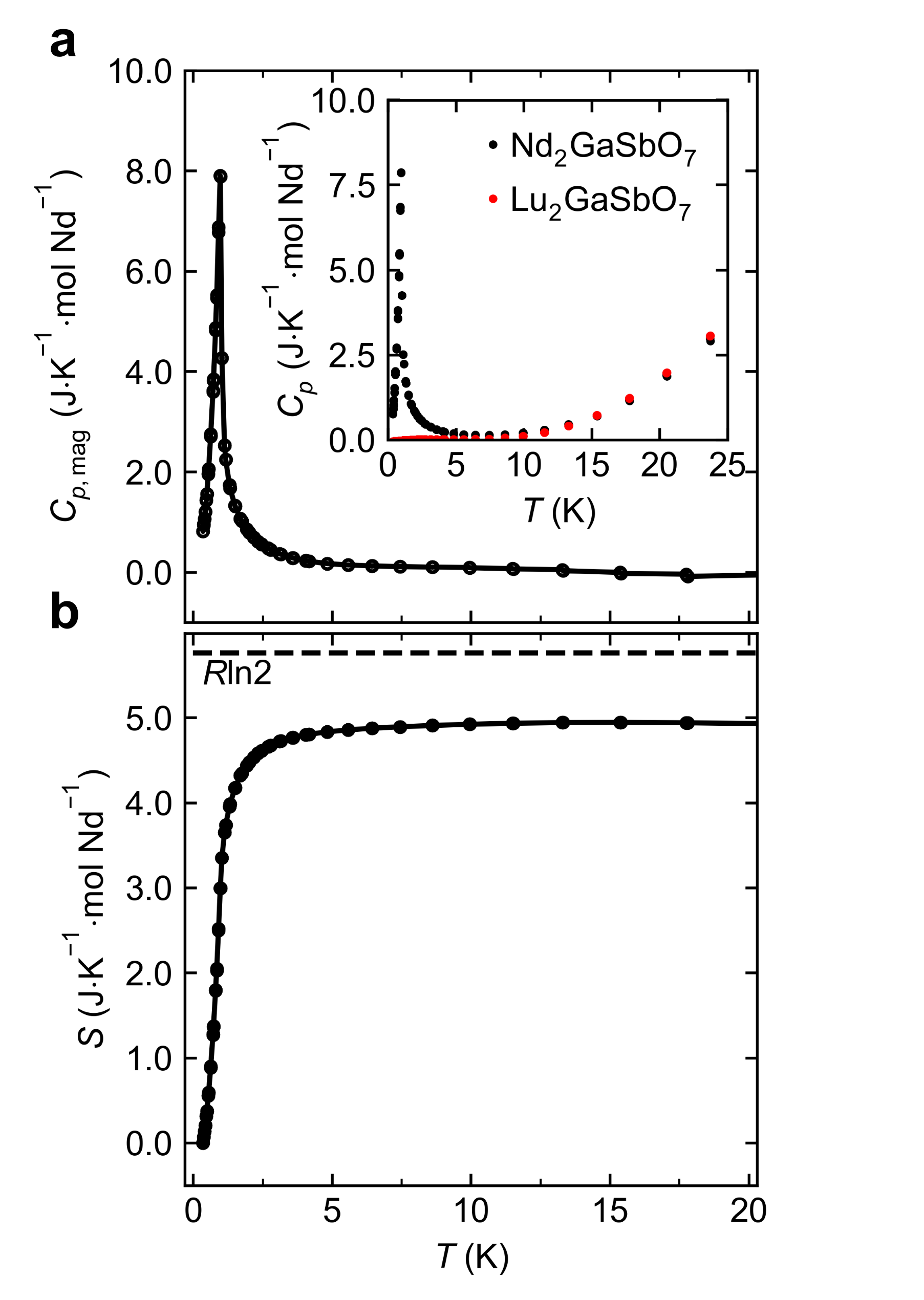}
	\caption{(a) Temperature dependence of the magnetic component of the heat capacity at 0~T for polycrystalline \ngso~revealing a $\lambda$-anomaly at 1.1~K, indicative of a transition to long range magnetic order. The magnetic component for \ngso~was isolated by subtracting off the normalized lattice contribution of the non-magnetic analog Lu$_{2}$GaSbO$_{7}$. The raw heat capacity data for both systems is shown in the inset. (b) Corresponding entropy release for \ngso~approaching $R\ln(2)$, as expected for a magnetic ground state doublet.}
    \label{figure:cp}
\end{figure}

As shown in Fig.~\ref{fig:XvTandMvH}(c), measuring the DC susceptibility down to $^{3}$He temperatures reveals the presence of a sharp maximum at $T_{\rm{N}}=1.1$~K. There is also a $\lambda$-anomaly at the same temperature in the heat capacity data shown in Fig.~\ref{figure:cp}(a). Taken together, these two features are indicative of a antiferromagnetic phase transition at $T_{\rm{N}}$. The magnetic transition temperature identified here is consistent with the value previously reported in an earlier study by Bl\"{o}te~\emph{et al} \cite{Blote1969}. Similarly to Nd$_{2}$Sn$_{2}$O$_{7}$, there is a clear absence of a broad hump at temperatures $T > T{\rm{_{N}}}$ that is usually attributed to the development of short range magnetic correlations in these geometrically frustrated magnets. Such behavior in \ngso~is in stark contrast to its ytterbium analog Yb$_{2}$GaSbO$_{7}$, where the $\lambda$-anomaly is completely absent and the heat capacity data is instead dominated by a broad hump centered at 2.3~K \cite{21_sarte}.

\begin{figure*}[t!]
	\centering
	\includegraphics[width=1\linewidth]{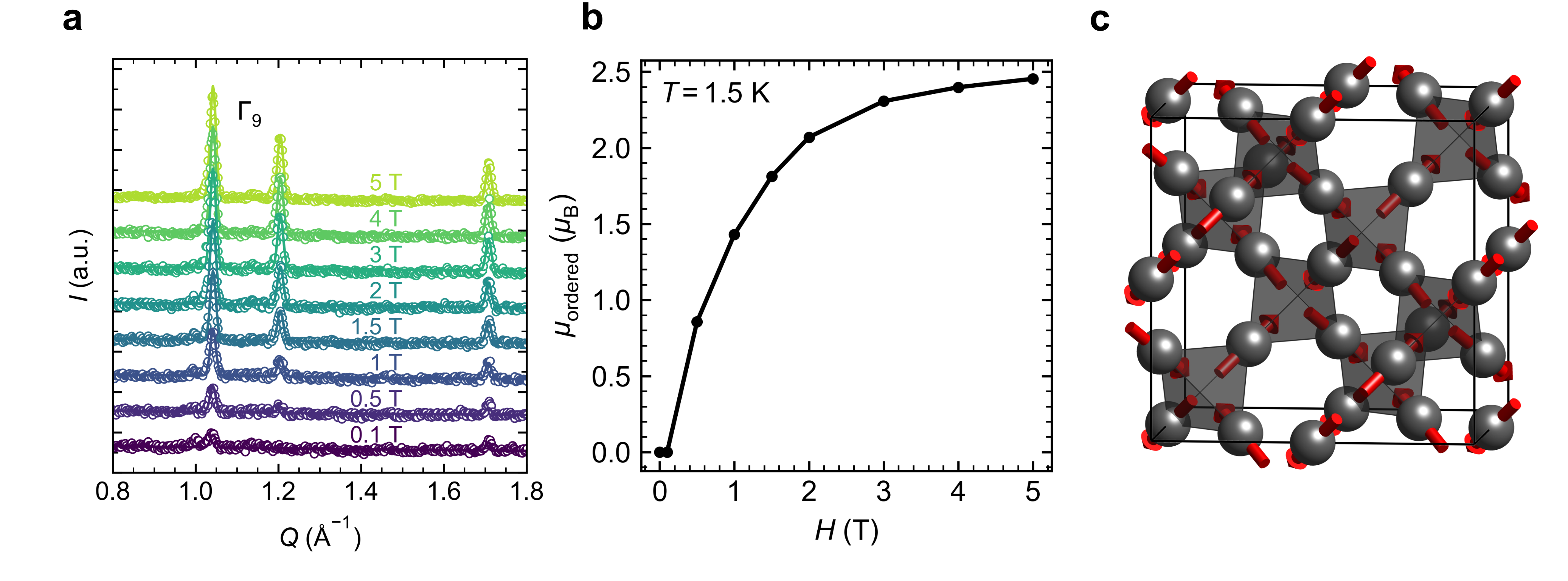}
	\caption{(a) Low-$Q$ neutron powder diffraction profiles and profile fits for polycrystalline \ngso~collected on HB-2A at 1.5~K under various applied fields between 0~T and 5~T. Additional scattering of magnetic origin appears for fields of 0.5~T and greater at the low $Q$ (111), (002), and (311) nuclear Bragg reflections, consistent with a $\Gamma_9$ magnetic structure. (b) Field dependence of the refined ordered moment from magnetic Rietveld refinements for \ngso~at 1.5~K highlighting a field-induced transition between $\mu_{0}H =$~0.1 and 0.5~$\rm{T}$. (c) Pictorial representation of the field-induced ferromagnetic $\Gamma_9$ ``ordered spin ice'' magnetic structure identified by magnetic Rietveld refinements for \ngso~at 1.5~K and fields of 0.5~T and greater. For the purposes of clarity, only Nd$^{3+}$ ions are shown.}
	\label{fig:hb2a-field}
\end{figure*}

From the second law of thermodynamics, the temperature dependence of the heat capacity $C_{p}$ determines the entropy change $\Delta S$ over a temperature range [$T_{1}$, $T_{2}$] by the following relation
\begin{equation}
    \Delta S =  \displaystyle\int_{T_{1}}^{T_{2}} \frac{C_{p}(T)}{T}dT.  
    \label{eq:secondlaw}
\end{equation}
By employing the Boltzmann definition of entropy $S=R\ln\Omega$, it becomes clear that the temperature dependence of C$_{p,{\rm{mag}}}$ in the low temperature limit allows for the direct experimental determination of the degeneracy of the magnetic ground state. As illustrated in Fig.~\ref{figure:cp}(b), the magnetic entropy released on warming from $T = 0.36~{\rm{K}}$ and $T = 22.5~{\rm{K}}$ is 0.89$R\ln{2}$, which is close to the value of $R\ln{2}$ expected for a Kramers doublet. The incomplete recovery of $R\ln2$ can likely be attributed to the presence of a significant amount of magnetic entropy remaining below 0.35~K, combined with a significant relative error ($\simeq$ 5\%) in the sample mass used for normalization.

\subsection{Neutron Powder Diffraction}

Having identified an antiferromagnetic transition at $T{\rm{_{N}}} =$~1.1~K, we employed neutron powder diffraction to characterize the magnetic moment configuration of the ordered state. As illustrated in Fig.~\ref{fig:hb2a}(a), a Rietveld refinement ($R_{p} =$~3.9$\%$, $R_{wp}$ = 5.3$\%$, $\chi^{2} =$~21) of the diffraction pattern collected at $T$~$>$~$T{\rm{_{N}}}$ confirmed that \ngso~crystallizes in the cubic $Fd\bar{3}m$ space group with a refined lattice parameter $a=10.33984(6)~{\rm{\AA}}$, representing the smallest lattice parameter reported for a Nd pyrochlore. The Rietveld refinement identified a small impurity peak at $Q =$~3.28~\AA$^{-1}$ corresponding to the fluorite-related Nd$_3$SbO$_7$ phase with a $<4\%$ volume fraction, a value below the detection limit of the preliminary standard laboratory X-ray diffraction measurement. Detailed results of the Rietveld refinement are provided in Table~\ref{table:rietveldresults}. No diffuse scattering, which could arise from short-range Ga$^{3+}$/Sb$^{5+}$ charge ordering and/or short range magnetic correlations, is visible in this data.
\begin{table}[tb!]
\caption{Crystallographic information at 10~K for \ngso, extracted from a Rietveld fit to the neutron powder diffraction data.}
\begin{ruledtabular}
\begin{tabular}{ccccc}
\rm{Atom}&
\rm{Wyckoff Position}&
$x,y,z$&
$U_{\rm{iso}}~({\rm{\AA}}^2)$\\
\colrule
        Nd  & $16c$ & 0,0,0     & 0.0071(7)\\
        \\[-1em]
        Ga/Sb  & $16d$ & $\frac{1}{2}$,$\frac{1}{2}$,$\frac{1}{2}$        & 0.0067(8)\\
        \\[-1em]
        O(1)  & $48f$ & 0.4229(1),$\frac{1}{8}$,$\frac{1}{8}$           & 0.0032(4)\\
        \\[-1em]
        O(2)  & $8b$  & $\frac{1}{8}$,$\frac{1}{8}$,$\frac{1}{8}$        & 0.014(1)\\
\end{tabular}
\end{ruledtabular}
\label{table:rietveldresults}
\end{table}\\
\indent
Upon cooling to $T$~$<$~$T{\rm{_{N}}}$, additional scattering intensity appears at the low $Q$ (220) and (311) Bragg reflections as shown in Fig.~\ref{fig:hb2a}(b); this is consistent with the bulk characterization results described above. All the magnetic Bragg peaks are resolution-limited, and can be indexed with a propagation vector $\mathbf{k} =$~(0 0 0), which is a common ordering wavevector for rare earth pyrochlores. In this case, the magnetic representation for the Nd$^{3+}$ ions can be reduced to four non-zero irreducible representations (IRs) given by $\Gamma_{\rm{mag}} = \Gamma^{1}_{3} + \Gamma^{2}_{5} + \Gamma^{2}_{7} + 2\Gamma^{3}_{9}$ in Kovalev's notation \cite{93_kovalev}, where $a\Gamma^{b}_{c}$ denotes the IR of the little group of $\mathbf{k} =$~(0 0 0) with an order $c$ and dimensionality $b$, occurring $a$ times.\\
\indent
Using the refined crystallographic parameters and scale factor at 10~K as initial estimates, a magnetic Rietveld fit to the diffraction pattern collected at 475~mK, also shown in Fig. \ref{fig:hb2a}(a), identified the magnetic structure as $\Gamma_{3}$ with an ordered moment $\mu\rm{_{ord}}$ = 1.59(5)~$\mu_{\rm{B}}/\rm{Nd^{3+}}$.  As illustrated in Fig. \ref{fig:hb2a}(c), $\Gamma_{3}$ corresponds to an AIAO magnetic structure, which is the ordered moment configuration previously identified in other Nd pyrochlores. More generally, this magnetic ground state is expected for the Ising antiferromagnets in this family of materials, as the moments are oriented along their local $\langle 111 \rangle$ axes. In contrast to all other IRs, such a non-coplanar antiferromagnetic structure is consistent with a lack of additional scattering intensity at both the (111) and (200) Bragg positions, as found for \ngso.\\
\indent
The ordered moment obtained from NPD is significantly reduced compared to the full-moment of the CEF ground state doublet given by $\mu_{\rm{CEF}} =$~2.37~$\mu_{\rm{B}}/\rm{Nd^{3+}}$, which may arise from an unsaturated moment (i.e. data collected at a temperature too close to the ordering transition) or may be indicative of strong quantum fluctuations within the CEF ground state manifold. While reduced ordered moments appear to be a hallmark of all Nd pyrochlores that order into the AIAO ground state, it is noteworthy that the systems argued to exhibit excited-state moment fragmentation have significantly smaller values (e.g., $0.80(5)~\mu_{\rm{B}}$ in Nd$_2$Zr$_2$O$_7$ \cite{Lhotel2015}).\\
\indent
As illustrated in Fig. \ref{fig:hb2a-field}(a), the application of an external magnetic field induces the appearance of additional scattering intensity centered at the low $Q$ (111), (002), and (311) nuclear Bragg reflections, in clear contrast to the $\Gamma_{3}$ behavior observed for the case of 0~T. Magnetic Rietveld refinements of diffraction patterns of polycrystalline \ngso~collected in various fields identified the magnetic structure as $\Gamma_9$. At 5~T, the refined magnetic moment $\mu_{\rm{ord}}=2.5(3)~\mu_{\rm{B}}$ is consistent with a fully-recovered CEF moment. The field-dependence of the refined magnetic moment presented in Fig. \ref{fig:hb2a-field}(b) confirmed a field-induced magnetic transition to the ferromagnetic ``ordered spin ice'' magnetic structure between $\mu_{o}H =$~0.1~T and 0.5~T, with a small canting angle away from the local $\langle 111 \rangle$ direction of $\sim$~3$^\circ$. A schematic of the ordered spin ice structure is shown in Fig. \ref{fig:hb2a-field}(c). This spin configuration has been identified in other Nd-based pyrochlores with magnetic transition metals on the $B$-site, both as the zero-field ground state (e.g., Nd$_2$Mo$_2$O$_7$) \cite{01_yasui} and induced by a magnetic field along the [001] direction (e.g., Nd$_2$Ir$_2$O$_7$) \cite{15_ueda}.\\
\begin{figure*}[htb!]
	\centering
	\includegraphics[width=0.9\linewidth]{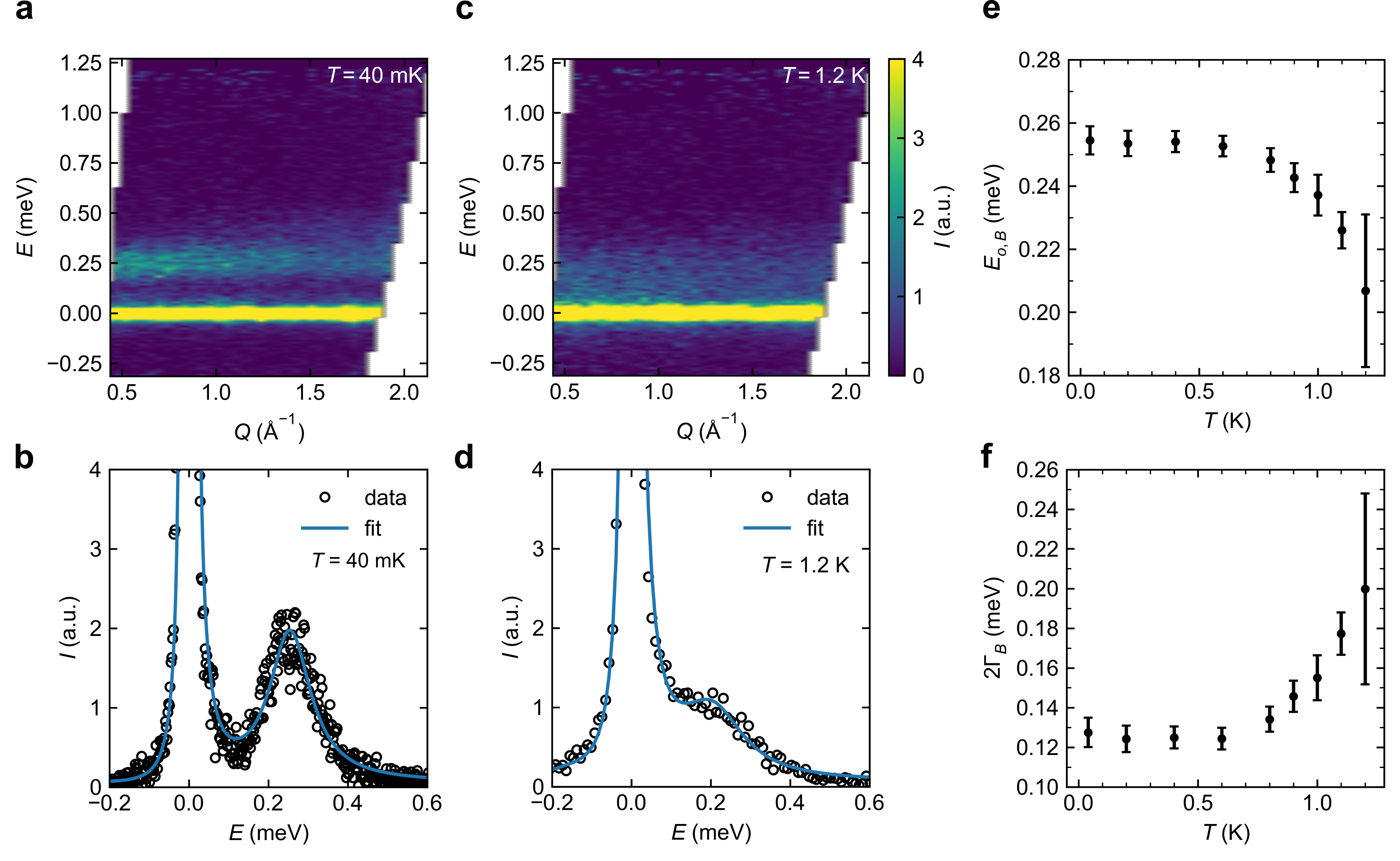}
	\caption{Color contour maps and corresponding $Q$-cuts (integration range of $Q =$~[0.5, 1.0]$~{\rm{\AA}}^{-1}$) of the neutron scattering intensity for polycrystalline \ngso~measured on IRIS using an $E_{f} =$~1.84~meV for (a),(b) $T < T_{\rm{N}}$ and (c),(d) $T > T_{\rm{N}}$. The constant $Q$-cuts are well-described by the lineshape given by Eq.~\ref{eq:lineshape}, as described in the main text. (e),(f) Temperature dependence of the lineshape fit parameters are consistent with the behavior of a spin-wave mode associated with long-range antiferromagnetic order.}
    \label{figure:iris}
\end{figure*}

\subsection{Low Energy Spin Dynamics}

To explore the possibility of moment fragmentation in \ngso, it is important to characterize the nature of the low-energy excitations since they play a key role in the realization of this phenomenon~\cite{Benton2016}. More specifically, flat and dispersive transverse spin wave modes should be visible that correspond to the divergence-free and divergence-full component of the fragmentation respectively, with the dispersive modes slightly higher in energy than the flat mode. Our inelastic neutron scattering measurements from the IRIS spectrometer at 40~mK, shown in Fig.~\ref{figure:iris}(a), identify the expected flat mode. Notably, the $Q$-dependence of the flat mode does not follow the characteristic behavior for spin ice scattering in a polycrystalline sample~\cite{Lefrancois2017}, which suggests that excited-state moment fragmentation is absent in \ngso.\\
\indent
A constant $Q$-cut through the flat mode with an integration range of $Q = [0.5, 1.0]$~$\rm{\AA}^{-1}$ is depicted in Fig.~\ref{figure:iris}(b), and it is well-described by the lineshape given by
\begin{equation}
\begin{split}
    S\left( E \right) =& \mathcal{A}_1 e^{-\frac{1}{2} \left( \tfrac{E-E_{o,A}}{\sigma_{A}} \right)^2}\\
    &+ \mathcal{A}_2\left( \tfrac{\Gamma_{A}}{\Gamma_{A}^2 + \left( E-E_{o,A} \right)^2}\right) \\
    &+ \mathcal{B} \left(n \left( E \right) + 1 \right) \left( \tfrac{\Gamma_B}{\Gamma_B^2 + \left( E-E_{o,B} \right)^2}\right)\\
    &- \mathcal{B} \left(n \left( E \right) + 1 \right) \left(\tfrac{\Gamma_B}{\Gamma_B^2 + \left( E + E_{o,B} \right)^2} \right) + \mathcal{C},
\end{split}
\label{eq:lineshape} 
\end{equation}
consisting of a linear combination of a Gaussian of width $\sigma_A$ and Lorentzian of width $\Gamma_A$ for the elastic peak with a center-of-mass $E_{o,A}$, and two Lorentzians of width $\Gamma_B$ for the inelastic peaks with centers-of-mass $\pm E_{o,B}$, while the instrumental background was approximated as a constant background $\mathcal{C}$. The inclusion of the Bose factor $n(E)$, combined with the anti-symmetrized linear combination of the two Lorentzians describing the inelastic feature, assures that the lineshape obeys detailed balance. Our fitted result yields a center-of-mass of $0.253(6)~{\rm{meV}}$ for the flat mode at 40~mK, as illustrated in Fig.~\ref{figure:iris}(b), which corresponds to a significantly larger value than reported for the moment fragmentation system Nd$_2$Zr$_2$O$_7$ (0.07~meV)~\cite{Petit2016} and the mixed $B$-site pyrochlore Nd$_2$ScNbO$_7$ (0.06~meV)~\cite{mauws2019}. The value for the center-of-mass shifts to lower energy transfers with increasing temperature and becomes gapless near $T{\rm{_{N}}}$, as shown in Figs.~\ref{figure:iris}(c) (d), and (e), while the width of the mode increases as the temperature approaches $T{\rm{_{N}}}$ from below (Fig.~\ref{figure:iris}(f)). The temperature-dependence of the spin gap matches expectations for a powder-averaged transverse spin wave mode that is associated with the AIAO antiferromagnetic order in this system \cite{Benton2016}. While the dispersive modes expected for this ordered state are not clearly visible here, this may be due to insufficient signal-to-noise in the polycrystalline experiment or an energy resolution that is too coarse to resolve them from the flat mode.

\subsection{Polarized Neutron Scattering}
\begin{figure}[htb!]
	\centering
	\includegraphics[width=1\linewidth]{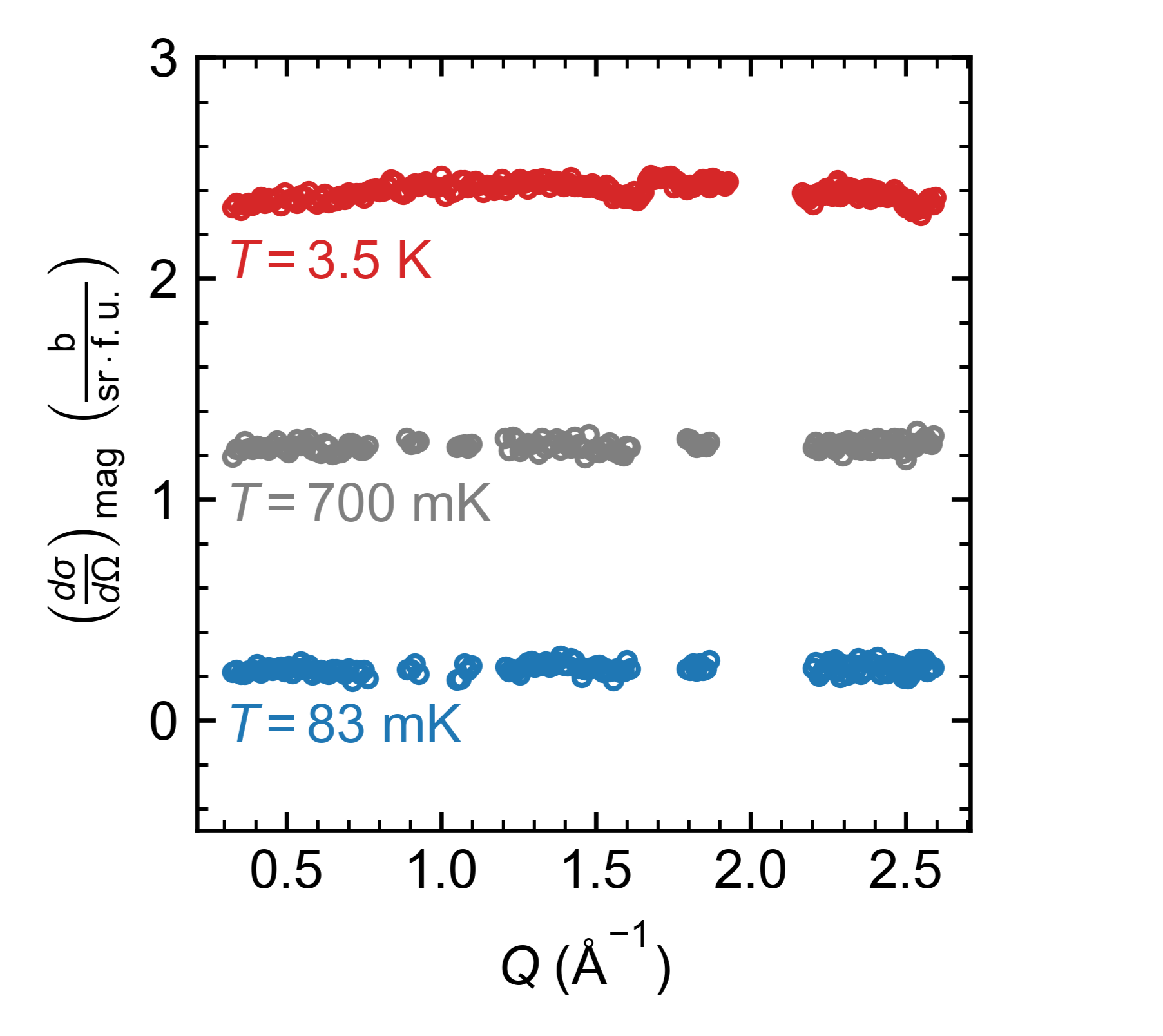}
	\caption{Comparison of the $Q$-dependence of the spin-polarized, energy-integrated ($\lambda =$~4.2~\AA) differential magnetic cross section for polycrystalline \ngso~at temperatures above and below $T\rm{_{N}}$ (Bragg peaks removed). For purposes of clarity, a vertical offset of $1~\frac{\rm{b}}{\rm{sr \cdot f.u.}}$ has been introduced for each successive temperature. 
	}
    \label{figure:diffusescattering}
\end{figure}

We collected polarized neutron scattering data on the DNS instrument to extend our search for the diffuse scattering associated with the divergence-free component of a fragmented state over a wider $Q$-range using an experimental set-up that facilitated a straightforward separation of the magnetic and non-magnetic scattering components via the 6 pt. xyz-polarization analysis method \cite{93_scharpf}. We note that this data is not energy resolved and effectively integrates the scattering over energy transfers up to 4.4~meV ($\lambda =$~4.2~\AA), so it will capture the flat inelastic mode measured on IRIS. As illustrated in Fig.~\ref{figure:diffusescattering}, our results confirm the absence of the spin ice correlations both above and below $T\rm{_{N}}$ expected for moment fragmentation systems \cite{Lefrancois2017}. In fact, no diffuse scattering is observed below $T\rm{_{N}}$, while the 3.5~K data consists of a broad feature centered at 1.25~\AA. This behavior is indicative of short-range magnetic correlations above $T\rm{_{N}}$ that disappear in the long-range ordered state, which is typical for conventional magnets.
\begin{figure}[htb!]
	\centering
	\includegraphics[width=1\linewidth]{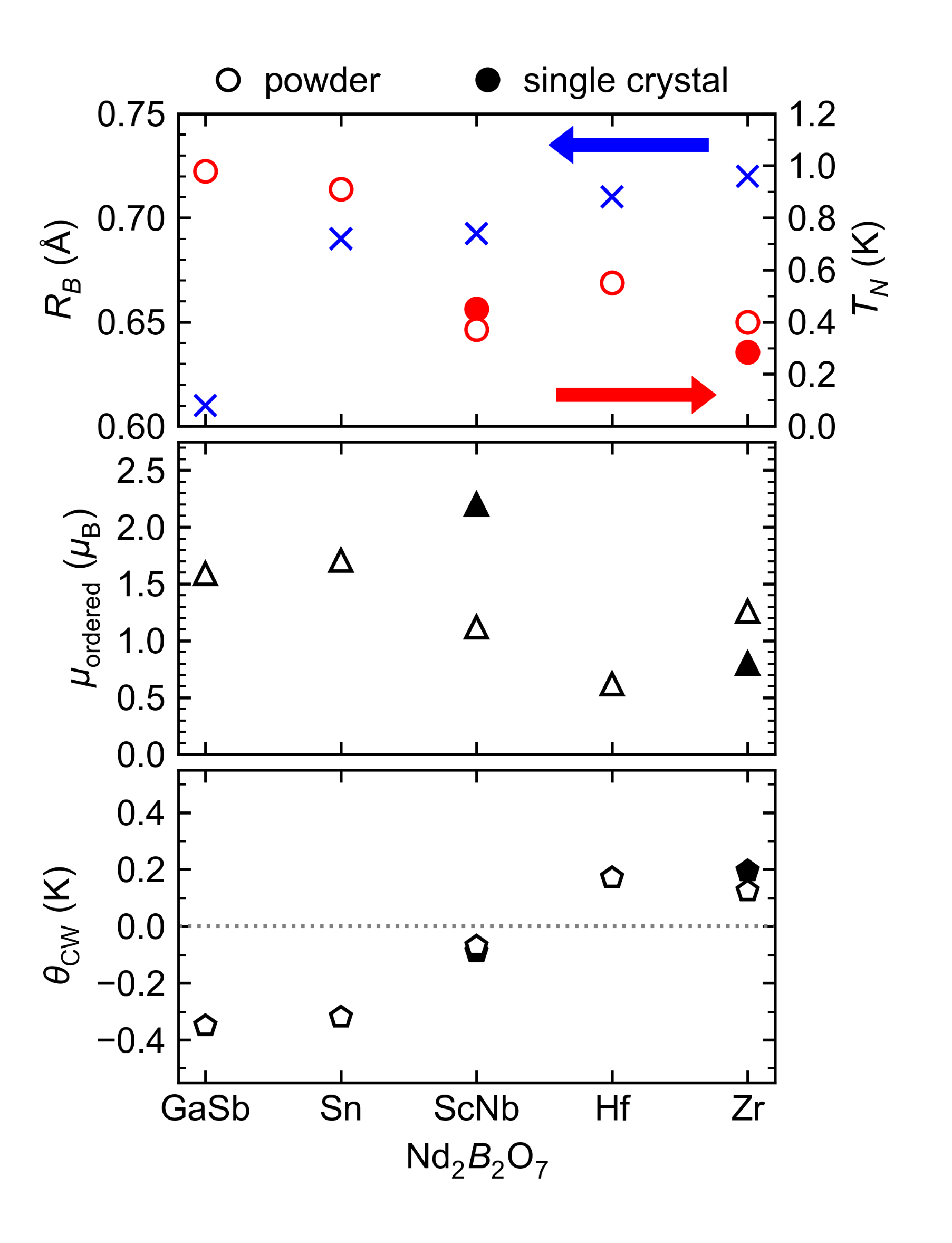}
	\caption{Comparison of select physical properties of Nd cubic pyrochlores for various $B$-site cations (arranged in order of increasing average ionic radius). Data was retrieved from \cite{Bertin2015} (Sn, powder), \cite{mauws2019} (ScNb, single crystal), \cite{21_scheie} (ScNb, powder), \cite{Anand2015} (Hf, powder), \cite{Lhotel2015} (Zr, crystal), and \cite{Xu2015} (Zr, powder). Open and filled markers denote powder and single crystal data, respectively.}
    \label{fig:Ndpyrochloretrends}
\end{figure}

\subsection{Discussion}

Upon a closer comparison of several key physical properties reported for all the Nd pyrochlores (Fig.~\ref{fig:Ndpyrochloretrends}) and those determined for \ngso~in the current study, two distinct material classes begin to emerge. The first class, where excited-state moment fragmentation is realized (e.g., Nd$_{2}$Zr$_{2}$O$_{7}$) or speculated (e.g., Nd$_{2}$Hf$_{2}$O$_{7}$), have larger ionic radii on the $B$-site, weak net ferromagnetic interactions ($\theta_{\rm{CW}}>0$), and significantly reduced magnetic transition temperatures and ordered moments. The second class, where moment fragmentation is absent (e.g., Nd$_{2}$Sn$_{2}$O$_{7}$), have smaller ionic radii on the $B$-site, weak net antiferromagnetic interactions ($\theta_{\rm{CW}}<0$), and larger magnetic transition temperatures and ordered moments. Nd$_{2}$ScNbO$_{7}$ represents an interesting crossover case due to its $\theta_{\rm{CW}}$ close to 0~K and the presence of rod-like diffuse scattering without evidence for pinch points both above and below the long-range magnetic ordering transition \cite{mauws2019, 21_scheie}.\\
\indent
From the current study, we find that Nd$_{2}$GaSbO$_{7}$ belongs to the second material class with no evidence for moment fragmentation. The reduction in the average $B$-site ionic radius results in a decrease in the lattice parameter $a$ and a corresponding enhancement of the Nd$^{3+}-$O$^{2-}-$Nd$^{3+}$ superexchange pathways, as shown by the concurrent increase in both $T_{\rm{N}}$ and the magnitude of $\theta_{\rm{CW}}$. Since the dominant net antiferromagnetic interactions provide no incentive for a magnetic moment to spontaneously flip and create magnetic monopoles, the realization of moment fragmentation in Nd$_{2}$GaSbO$_{7}$ is precluded and instead replaced by an AIAO antiferromagnetic ground state possessing an enhanced ordered moment. The general trends presented in Fig.~\ref{fig:Ndpyrochloretrends} clearly establish that chemical pressure plays a critical role in driving fragmentation phenomena in the Nd pyrochlores, while introducing significant $B$-site disorder has minimal impact on altering the collective magnetic properties of this material family. Notably, the $B$-site disorder does not generate a spin glass ground state in \ngso~and instead AIAO magnetic order is preserved.

\section{Conclusions}

In summary, our bulk characterization and neutron scattering measurements have identified a clear absence of excited-state moment fragmentation in polycrystalline \ngso. The small lattice constant, relative to other Nd pyrochlores, seems to enhance the antiferromagnetic exchange interactions required to generate conventional all-in-all-out magnetic order in pyrochlore systems with Ising anisotropy. Our work shows that chemical pressure plays an important role in the realization (or absence) of excited-state moment fragmentation in the Nd pyrochlore family. On the other hand, $B$-site disorder seems to have minimal impact on the collective magnetic properties of the Nd pyrochlores. Nd$_{2}$ScNbO$_{7}$, with its average $B$-site ionic radius of intermediate size, is an interesting crossover case that warrants future investigation.

\section{Acknowledgments}

The authors acknowledge fruitful conversations with Cole~Mauws, Brenden~Ortiz, Mitchell Bordelon, Kasey~Camacho, Rachel~Camacho, Cassandra~Schwenk, Yared~Wolde-Mariam, Andrea~Reyes, and Casandra~Gomez~Alvarado. S.J.G. acknowledges financial support from the National Science Foundation Graduate Research Fellowship under Grant No. 1650114. P.M.S. acknowledges financial support from the University of California, Santa Barbara through the Elings Fellowship. P.M.S. acknowledges additional financial support from the CCSF, RSC, ERC, and the University of Edinburgh through the GRS and PCDS. B.A.G. acknowledges financial support from the University of California Leadership Excellence through Advanced Degrees (UC LEADS) Fellowship. C.R.W. acknowledges financial support from the CRC (Tier~II) program, the Leverhulme Trust, CIFAR, CFI and NSERC. S.D.W. acknowledges financial support from the US Department of Energy (DOE), Office of Basic Energy Sciences, Division of Materials Sciences and Engineering under Grant No. DE-SC0017752. J.P.A., K.H.H., and E.J.P. acknowledge financial support from the ERC, the STFC and the RSC. This research was undertaken thanks in part to funding from the Max Planck-UBC-UTokyo Centre for Quantum Materials and the Canada First Research Excellence Fund, Quantum Materials and Future Technologies Program. A portion of this research used resources at the Spallation Neutron Source and High Flux Isotope Reactor, which are DOE Office of Science User Facilities operated by Oak Ridge National Laboratory. Inelastic neutron scattering experiments on IRIS at the ISIS Neutron and Muon Source were supported by the beamtime allocation RB1720442 (\href{https://doi.org/10.5286/ISIS.E.RB1720442}{https://doi.org/10.5286/ISIS.E.RB1720442}) to P.M.S. and J.P.A. from the Science and Technology Facilities Council (STFC). The authors gratefully acknowledge support via the UC Santa Barbara NSF Quantum Foundry funded via the Q-AMASE-i program under award DMR-1906325. The authors would also like to thank the Carnegie Trust for the Universities of Scotland for providing facilities and equipment for chemical synthesis. Finally, the authors gratefully acknowledge technical support provided by Heinrich Kolb and financial support provided by the J\"{u}lich Centre for Neutron Science (JCNS) to perform the neutron scattering measurements at the Heinz Maier-Leibnitz Zentrum (MLZ), Garching, Germany.

\renewcommand{\figurename}{Figure S}
\renewcommand{\topfraction}{0.85}
\renewcommand{\bottomfraction}{0.85}
\renewcommand{\textfraction}{0.1}
\renewcommand{\floatpagefraction}{0.75}


%

\end{document}